\documentclass[12pt,preprint]{aastex}

\newcommand{\um}{\,$\mu$m}

\def\aj{{AJ}}

\def\apj{{ApJ}}
\def\apjl{{ApJL}}
\def\apjs{{ApJS}}

\def\aap{{A\&A}}

\def\mnras{{MNRAS}}

\def\qjras{{QJRAS}}

\shorttitle{
Cold Dust 
in Nearby Ellipticals}
\shortauthors{Leeuw et al.}

\begin{document}

\title{
Observations of Cold Dust in Nearby Elliptical Galaxies}

\author{Lerothodi L. Leeuw,
\altaffilmark{1}
\altaffiltext{1}{
Ritter Astrophysical Research Center,
University of Toledo, Mail Stop 113, 
2801West Bancroft Street, Toledo, OH 43624;
leeuw@astro1.panet.utoledo.edu.}
Anne E. Sansom,
\altaffilmark{2}
\altaffiltext{2}{
Centre for Astrophysics, University of Central Lancashire,
Preston PR1 2HE, United Kingdom; aesansom@uclan.ac.uk.}
E. Ian Robson,
\altaffilmark{3}
\altaffiltext{3}{
Astronomy Technology Centre, Royal Observatory,
Blackford Hill, 
Edinburgh EH9 3HJ, United Kingdom; eir@roe.ac.uk.}
Martin Haas,
\altaffilmark{4}
\altaffiltext{4}{
Astronomisches Institut, Ruhr-Universit\"at Bochum, 44780 
Bochum, Germany; haas@astro.ruhr-uni-bochum.de.} 
\\
and
Nario Kuno
\altaffilmark{5}
\altaffiltext{5}{
Nobeyama Radio Observatory, Minamimaki, Minamisaku, Nagano 
384-1305, Japan; kuno@nro.nao.ac.jp.}
}

\begin{abstract}
Spectral energy distribution (SED) analyses, that include new
millimeter (mm) to far-infrared (FIR) observations, obtained with
continuum instruments on the Nobeyama and James Clerk Maxwell Telescopes and
 {\it Infrared Space Observatory} ({\it ISO}), are presented for
seven, nearby ($<45$\,Mpc) FIR-bright elliptical galaxies.   
These are analyzed together with archival FIR and short-wave radio data
obtained from the NASA/IPAC Extragalactic Database (NED).  The radio
to infrared SEDs are best fit by power-law plus greybody models of
dust residing in the 
central galactic regions within 2.4\,kpc diameter and with temperatures
between $\sim 21$ and 28\,K, emissivity index $\simeq 2$, and masses
from $\sim 1.6$ to $19 \times 10^5 {\rm M}_{\odot}$. 
The emissivity index is consistent with
dust constituting 
amorphous silicate and carbonaceous grains previously  
modeled for stellar-heated dust observed in the Galaxy and
other nearby extragalactic sources.  Using 
updated dust absorption
coefficients for this type of dust, dust masses are estimated that
are similar to those determined from earlier FIR data alone, even though the
latter 
results implied hotter dust temperatures.
Fluxes and masses that are consistent with the new FIR and submm data
are estimated
for dust cooler than 20\,K within the central
galactic regions.  Tighter physical constraints for such   
cold, diffuse dust (if it exists) with {\it low} surface brightness, 
will need sensitive FIR-submm observations
with the {\it Spitzer Space Telescope}, SCUBA2, or ALMA.

\end{abstract}


%\keywords{galaxies : dust -- galaxies: individual ( NGC\,2986,
%NGC\,3156, NGC\,3962, NGC\,4374, NGC\,4697, NGC\,5353/NGC\,5354) --
%galaxies: ellipticals}

\section{Introduction}
\label{sec:intro_es}

Optical-absorption patches, lanes, and filaments of dust have been
seen in 50 to 80\% of nearby bright elliptical galaxies
\citep[e.g.][]{von95}.  Observations at a range of other 
wavelengths have also revealed unexpected amounts of gas and dust in
these galaxies \citep[e.g.][]{rob91, gou94_3}.  In some objects, the
dust masses estimated from optical extinction studies are
%significantly
a magnitude lower than masses implied by the $IRAS$ 
%galaxies' 
far-infrared (FIR)
%far-infrared
 fluxes, suggesting that ellipticals may contain diffusely
distributed dust not detected or properly accounted for in optical
observations \citep[e.g.][]{gou95}.  
The dust would add another degeneracy to age-metallicity degeneracies,
further complicating the interpretations of broadband colors
\citep[e.g.][]{sil96a}.   Attenuation by the dust would also produce
effects similar to the expected kinematical signature of a dark matter halo,
%such that dust attenuation may form an
and thus provide an alternative explanation for the
usual stellar kinematical evidence for dark matter haloes around
elliptical galaxies \citep[e.g.][]{bae02}.  Therefore, constraining
the physical properties of the dust 
%in these galaxies
 is very crucial.
 
The $IRAS$ FIR 
%observations were made with 
beams were 
%greater than
$> 1'$ and sampled only the spectrum of dust 
%warmer than 
at temperatures $> 25$\,K,
% that peaks shortward of 100\,\um,
providing 
%rather 
poor constraints on the dust distribution
and emission from dust 
%cooler than 
$< 25$\,K. 
%constrain the 
%cold 
%dust properties in ellipticals were made by
%In early submm 
%studies to 
%observations with 
Using the previous generation submm
detector, UKT14 \citep{dun90}, 
\citet{kna91}
did not detect any submm dust emission
%above that expected from an extrapolation of the radio-synchrotron
%emission 
from their sample 
of radio galaxies, that included many $IRAS$-detected
ellipticals.  
%Following galactic dust estimates, 
%For their 
%radio 
%galaxies, \citet{kna91}  
They assumed dust
temperatures of $\sim$ 18\,K, as 
estimated 
for Galactic dust \citep[e.g.][]{rea95_et}; and, boosted
by the lower 
%assumption of 
dust temperatures, they inferred large
dust contents of $\sim 10^6-10^8 {\rm M}_{\odot}$, i.e. 
%of the order of 
as found in luminous spiral
galaxies.  Millimeter and submm continuum 
%dust 
emission was
subsequently detected from a sample of early-type galaxies by
\citet{fic93} (also using UKT14).  Results from these data showed that
dust masses in early-type galaxies were an order of magnitude
lower than seen in spirals.  
%that strongly suggested ellipticals
%may habor modest cold dust close to the order 
%Their lower assumption of the dust temperatures leads to significantly
%larger dust mass determinations than those of \citet{gou95}.
%A more recent analysis of ellipticals compilation of d
In a recent review of cold gas in elliptical galaxies, \citet{kna99}
concluded that field ellipticals contain about 0.01 to 0.1 of the
interstellar matter content of spiral galaxies of similar luminosity.  
%and support a low level of star formation.  
%However, 
Because of the dust's
relative paucity, detecting its total content in ellipticals is much harder.
%the total dust in ellipticals is much harder to detect.

A re-examination of the $IRAS$ detection rate and fluxes by
\citet{bre98} showed that most $IRAS$ detections of early-type galaxies
were near the $3\sigma$ level, with only about 12\% of the galaxies in
their sample having been detected above the 98\% confidence level.
This showed that although dust might be present in a number of
early-type galaxies, a record of 
%Those authors' aim was not to demonstrate that only a few early-type galaxies
%are FIR emitters.  Nevertheless, their results show that
%are a reminder
%rather to determine the galaxies for which 
good FIR fluxes, that were crucial for constraining dust
properties in ellipticals, existed for only a few early-type galaxies.
Recent, extensive work by \citet[][]{tem04} added 
%should add improve on the
$ISO$ archival data for 39 giant elliptical galaxies to the FIR data record of
ellipticals and is expected to be complemented by new FIR data from the
%currently commissioned
recently launched {\it Spitzer Space Telescope}. 

This paper extends the previous FIR to mm studies of early-type
galaxies, exploiting newer, sensitive instruments, by exploring whether
2\,mm to 450\,$\mu$m data obtained with
NOBA and SCUBA [which is 100 times more sensitive than UKT14 \citep{hol99_et}], can reveal any dust component
$<20$\,K and constrain the total content in selected nearby field ellipticals. 
%, as seen in other nearby extragalactic sources.  
The mm to submm measurements from NOBA and SCUBA, which are obtained respectively at the Nobeyama\footnote{The Nobeyama Telescope is
operated by 
%the National Astronomical Observatory of Japan.} 
the Nobeyama Radio Observatory (NRO), which is a branch of the National
    Astronomical Observatory of Japan.}
%, an inter-university research institute operated by
%    the Ministry of Education, Culture, Sports, Science and Technology. 
and 
%Telescope and SCUBA on the 
James Clerk Maxwell Telescopes (JCMT),\footnote{The JCMT is operated by the Joint Astronomy Centre on behalf of the United Kingdom Particle
Physics and Astronomy Research Council (PPARC), the Netherlands Organisation for Scientific
Research, and the National Research Council of Canada.} are used together with 
%the FIR 
archival 60
and 100\,$\mu$m $IRAS$-NED\footnote{
%The $IRAS$ and radio 5\,GHz fluxes were obtained from NED t
The NASA/IPAC Extragalactic Database (NED) is
operated by the Jet Propulsion Laboratory, California Institute of
Technology, under contract to the National Aeronautics and Space
Administration.}
and new 60 to 200\,$\mu$m $ISO$ data,\footnote{Based on observations with the Infrared Space Observatory
         ISO, an ESA project with instruments
         funded by ESA Member States (especially the PI countries: France,
         Germany, the Netherlands and the United Kingdom) and
         with the participation of ISAS and NASA.}
 with the aim of 
tightening 
%the 
physical properties
of cold 
%to cool 
dust that emits in these 
%mm to FIR 
wavebands.  The thermal
emission fluxes from the mm to FIR continuum are analyzed together 
with some archival radio data, obtained from NED, to account for any non-thermal, high-frequency radio
synchrotron contribution to the cold dust spectrum.   
%Furthermore,
Some $IRAS$ data were re-processed using HIRES- and SCANPI routines
at
% courtesy of 
the NASA/IPAC Center and used to check the 
%$IRAS$-NED 
data for any FIR source confusion or contamination.

%%\section{Millimeter-Submm Observations and FIR
%%Data}\label{sec:obs_es} 

\section{Source Selection, Far-infrared Data, and Submm-to-mm
%Millimeter
Observations}\label{sec:obs_es}

The elliptical galaxies selected for submm-to-mm observations had
$IRAS$-detected 60 and 100\,$\mu$m emission that was thought to
indicate excess dust over their optically determined dust masses
\citep[e.g.][]{gou94_3, sil96a}.   
%With the goal of directly detecting and investigating the reality of
%this excess dust in new FIR data and sub-to-mm observations, 
In order to maximize the detection of expected, low-level fluxes
% detections 
and with the aim of spatially resolving them, the targets are relatively strong
FIR emitters
($>100$\,mJy) and nearby ($<45$\,Mpc).  To avoid confusing any non-thermal
flux contribution to the detected thermal emission, the galaxies,
apart from NGC\,4374, are low-radio emitters.  They include
ellipticals that had inferred dust masses
greater than $\sim 10^5 {\rm M}_{\odot}$
in samples by \citet[][]{kna91, gou95}; and \citet[][]{bre98}.

ISOPHOT photometry observations from 60 to 200\,$\mu$m \citep{lem96_et}
%(Lemke et al. 1996)
 of the five galaxies NGC\,3962, NGC\,4374, NGC\,4697, and
NGC\,5353/NGC\,5354
% NGC\,2986, NGC\,3156, 
were obtained from the $ISO$ Data Archive \citep{kes00}.  
%(Kessler et al. 2000).
They were reduced using the PHOT Interactive Analysis tool (PIA V9.1)
together with the calibration data set V7.0, providing a photometric 
accuracy of better than 30\% for faint sources \citep{lau99}.
%(Laureijs $\&$ Klaas 1999).

NGC\,3962, NGC\,4374 and NGC\,4697 were observed in chopped mode (ISOPHOT
AOT 22) and NGC\,5353/5354 in sparse mapping (ISOPHOT AOT 37/39).
The data reduction included correction for the electronics'
non-linearity, removal of data sections contaminated by cosmic
particle events (also known as deglitching), and
correction for signal dependence on the reset interval time.  For the
chopped observations, the data reduction also included correction for
transient effects using the so-called ``pattern-analysis'' method.
% \citep[e.g.][]{haa00}.
%(e.g. Haas et al. 2000).
  As a cross check on 
%the ``pattern-analysis'' 
this technique, a
Fourier-analysis method 
was also applied as described in detail by \citet[][]{haa00},
%Haas et al. 2000.
%The two
%methods 
yielding similar results to within 10\%. 
Further reduction included the calibration of the detector
responsivity and its changes, performed using associated measurements
of the thermal fine calibration source on board. 

All $ISO$ source fluxes were corrected
for aperture size assuming an unresolved point
source, since on the ISOPHOT C100 array (which has 3 $\times$ 3 pixels 
of diameters 46$\arcsec$) no indication for any significant, extended
60-to-100\,$\mu$m  flux was found beyond the pixel resolution.
%diameter of 46$\arcsec$.  
The photometry in the 60-to-100\,$\mu$m wavebands was
derived from the 46$\arcsec$, central-C100 pixel pointed at the source.
For the chopped 170 and 200\,$\mu$m observations, where the sources were
centered on the 2 $\times$ 2 pixel array, the photometry was derived
from the whole array (180$\arcsec$), as in that case the source flux
was measured by all four pixels (see, e.g., top-left panel of
Figure~\ref{fig:arrays}).  Slightly different
flux extraction was required for the 180\,$\mu$m observations of
the companion galaxies NGC\,5353 and NGC\,5354, because both objects were
detected by this array simultaneously (see details below). 
%described below).
% and not resolved from each other
%(see details below). 

The top panels in Figure~\ref{fig:arrays} show that the galaxies NGC\,5353 and NGC\,5354
are only about 70$\arcsec$ apart, with the latter being in the north.
This is sufficiently close for 
the two galaxies to be covered by the ISOPHOT C100 and
C200 arrays, that have $3 \times 3$ and $2 \times 2$ pixels of size
46$\arcsec$ and
92$\arcsec$ each, respectively, and to allow the galaxies to be resolved on
the C100 array.  
%Therefore, 
As shown on the left panels in Figure~\ref{fig:arrays}, NGC\,5353 and NGC\,5354
were respectively observed exactly with the center and corner
pixels of the
%$3 \times 3$ pixel, 
C100 array, ensuring that the
galaxies would be clearly resolved from each other on this array.
Those C100 pixels which are not centered on the two sources show some 
low-level flux (bottom-left histogram of Figure~\ref{fig:arrays}),
which stems 
from the wings of the point spread function, primarily of the
central source.  The same applies for the 
observations at 60\,$\mu$m.
The right panels of Figure~\ref{fig:arrays} show that at 180\,$\mu$m
NGC\,5353 is centered on the 2 $\times$ 2 pixels of C200
array, so that its flux is distributed over all four pixels,
and NGC\,5354 is located on the corner pixel containing most
of its flux.  In order to separately extract the 180\,$\mu$m fluxes of
the two galaxies on the 
C200 array, two point sources were
fitted to the measured fluxes -- one was centered on the array and the other
located almost exactly on a corner pixel, corresponding to the
C100 array positions of NGC\,5353 and NGC\,5354 respectively.
%HIRES-reprocessed-$IRAS$ data also show that the companion galaxies
%NGC\,5353/NGC\,5354 are both FIR-bright clearly resolved from each
%other in the $IRAS$ 60 and 100\um\ bands.  Therefore, 
This paper exploits the new $ISO$ data of NGC\,5354 and NGC\,5353 and
deals with these companion galaxies as individual FIR sources.  For
the first time ever (see details about the corresponding $IRAS$
measurements below), the
two galaxies are resolved from each other at 60 and 100\,\um, and their
fluxes are separately extracted at 180\,\um.  It is noted that
\citet[][]{tem04} did not resolve the pair NGC\,5354 and NGC\,5353 
%or present to the pair 
in their recent, extensive analysis of the same $ISO$ data as presented here.

Photometric observations in the submm were obtained with SCUBA on
various nights from 2001 January to 2002 February.  The camera was
simultaneously operated with the central bolometers of 
the short-wave array at 450\,$\mu$m and the long-wave array
at 850\,$\mu$m. 
%, with beam-widths of $8''.5$ and $14''.5$ (FWHM) respectively.  
The photometry employed a 9-point jiggle pattern in a $3 \times 3$
grid of $2''$, and standard SCUBA data reduction was
undertaken \citep[e.g.][]{lee00}.  The fluxes were calibrated using
instrumental gains 
%that were 
determined from photometry observations
of planets and, when planets were not available, the JCMT
secondary calibrators \citep{san98, jen02}.

On 2002 March 4, the ellipticals NGC\,4374, NGC\,4697,
and NGC\,5353 were scan-mapped 
%observed 
at 2\,mm with the NOBA bolometer array
%in scan mapping mode.
attached to the Nobeyama--45-m Telescope.   During the telescope allocation
for this study, only three objects were observed
because there was only a limited 
%by a {\em short} at that time at the Nobeyama Radio Observatory
spell of good weather.  The typical sky opacity (tau) during the
observation was 0.12, measured using skydips. 
Exposure times were $\sim$90\,minutes per galaxy, raster-scanning across
each target to obtain maps of $1 \times 1$\,arcminutes. 
Primary flux calibration observations were obtained on Saturn and
secondary calibration on 3C273 and
3C279.  The expected flux for Saturn was obtained from the JCMT
planetary flux program and those for 3C273 and 3C279 were
boot-strapped from JCMT monitoring observations.  There was no
indication of extended emission beyond the NOBA-$12''$ beam (FWHM) at
2\,mm.  The resultant FIR to mm flux estimates and upper limits are presented
in Table~\ref{tab:zero}.  

There is a discrepancy between the $IRAS$ and $ISO$ fluxes of NGC\,5353
and NGC\,5354; this is because, on $IRAS$ scans, NGC\,5354 was not resolved from
NGC\,5353, and only the combined flux of the galaxy pair was listed in
the $IRAS$ point source catalog (PSC).  Compared with the combined
$ISO$ fluxes of the galaxies, 
%NGC\,5353 and NGC\,5354, 
the $IRAS$ PSC flux is much
smaller, i.e., only about 50\%.  The baseline determined
by the $IRAS$ PSC algorithm generally uses an unresolved point source filter, so that
for a moderately extended source like the pair NGC\,5353/5354 the
baseline is set at the level of the extended emission, i.e., too high,
resulting in too a low source flux.
A similar trend of too low $IRAS$ PSC fluxes was found for 12 moderately
resolved spiral galaxies \citep[][their section 4.1.3]{ben02}.
%(Bendo et al. 2002, their section 4.1.3).
%In addition, we examined 
An examination of the SCANPI-processed-$IRAS$ scans around NGC\,5353 shows
that the low $IRAS$ fluxes might be caused by a poor baseline
determination on the pair NGC\,5353/5354, plus some extended flux
contribution from another 4$\arcmin$ nearby source.
A visual inspection of the scans and baselines for the pair
gives F(60\,$\mu$m) = 0.45\,Jy and F(100\,$\mu$m) = 1.9\,Jy, which
comes close to the $ISO$ values of F(60\,$\mu$m) = 0.6\,Jy and
F(90\,$\mu$m) = 1.95\,Jy. This leads to a conclusion that
the 60 and 100\,$\mu$m $ISO$ fluxes of NGC\,5353 and NGC\,5354 in this
paper are correct and the newly determined fluxes are used in the spectral
energy distribution (SED) analyses below.
%The SEDs in Section who 
It is interesting that despite the decomposition of the
180$\mu$m fluxes, the SEDs of NGC\,5353 and NGC\,5354
are not much different from those of the other sources presented in
this paper (see
Section~\ref{sec:sed_es} below).

\section{Beam Sizes and Source Fluxes and Extents}\label{sec:beam_es}

The fluxes presented here (see Table~\ref{tab:zero}) come from single-pointing
observations obtained with beams that vary in size.  The $IRAS$ 60 and
100\,$\mu$m are $\sim 120''-240''$, $ISO$
170 to 200\,$\mu$m $180''$ (except $92''$ for NGC\,5353 and NGC\,5354), $ISO$ 60 to 100\,$\mu$m $\sim 46''$, SCUBA
850\,$\mu$m $\sim 16''$, NOBA 2\,mm $\sim 12''$, and SCUBA 450\,$\mu$m
$\sim 9''$ (FWHM in all cases, except the $180''$ for the $ISO$
170 to 200\,$\mu$m is $2 \times$ FWHM).  Therefore,
if the FIR to submm emission from the targets is
as extended as the galaxies' 
%stellar dimensions, i.e. $1'.6$ to $7'.6$
effective radii, i.e. $14''$ to $72''$
(see Table~\ref{tab:one}),
then for each object, the fluxes from the respective instruments will come
from different galactic regions corresponding to the sampling beam sizes.  

%Well aware of the above, 
Mindful of this and for reasons noted below, the SED analyses here
assume that all the 
%$IRAS$, $ISO$, NOBA and SCUBA 
fluxes come from predominantly
unresolved, central regions of the galaxies, whose extent is less 
%than the regions sampled by 
than the employed mm to $ISO$-100\,$\mu$m beams, i.e. $<8''$ to
$<46''$, or roughly $<1$\,kpc to $<$6\,kpc galactic scales.  
%Recent HST observations 
%\citep[e.g.][]{von95} have shown that 
As noted earlier, at least 50\% of nearby bright
ellipticals have dust features that manifest absorbing optical light
within areas of such nuclear regions.  
Some ellipticals also show a compact nucleus associated with radio
emission in some of these galaxies \citep[e.g.][]{bow97}. 
The submm emission from these cores is expected to be the high frequency
extrapolation of synchrotron radiation that is responsible
for the radio source.   \citet[][]{lee02} showed that this emission
can be 
%at scales 
comparable to the total submm
flux from low level dust emission, and therefore needs to be
accounted for when analyzing any possible submm re-radiation from 
%the
dust, especially in unresolved sources with active nuclei.

The assumption that the FIR to submm emission comes from a
dominant unresolved region in the presented observations is made
following NOBA-, SCUBA- and ISOPHOT-mapping results of 
%two the 
three of the galaxies studied here, i.e. NGC\,4374, NGC\,5353,
and NGC\,5354.
First, the SCUBA mapping
observations by \citet[][]{lee00} showed that the 850\,$\mu$m
continuum from this object was unresolved, constraining the emitting
source by the SCUBA-850 resolution to $< 16''$.  Second, the NOBA mapping
results of NGC\,4374 and NGC\,5353 constrain the 2\,mm-emission from
these galaxies by the NOBA resolution to $< 12''$. 
Similarly, the ISOPHOT mapping results constrain the 100\,$\mu$m-emission
from NGC\,5353 and NGC\,5354 to $< 46''$. 
Consistent with the
above assumption, the parameter results of the SED model fitting
require that the emitting-greybody sources in all the studied
ellipticals are less than about $9''$ (see below).  This is true for
single-temperature-greybody sources with the standard dust emissivity
indices from 1 to 2.  However, in the case the SED model fitting is
computed for two-(or-more)-temperature-greybody sources \citep[][]{lee02},
where one of the temperatures is considerably lower than 
%temperatures
results presented in this paper, larger emitting regions 
for the coldest dust could be implied.     

%Since the 

Removing the assumption that the emitting regions detected here are
unresolved could have some 
consequence on the fitting and
derived parameters.  
In particular, if the detected FIR to mm emission 
%is resolved, 
is extended, the
larger FIR beams could include emission from outside the very central
regions ($\sim 8''$ to $16''$) sampled by the mm and submm beams and
%This would 
imply that the mm and submm fluxes are
undersampled relative to the FIR data in the SED fits.
The SED fitting to the ``undersampled'', lower submm fluxes ($F_{\nu}$
or $S_{\nu}$ below)
would 
%produce
(a) require higher emissivity index ($\beta$) values 
%than would be needed 
to match the slope
%for higher submm fluxes 
at the Raleigh Jeans tail, (b) imply smaller emitting region
or scaling factor ($\Omega$) values (from Equation~\ref{eqn:pb},
$F_{\nu} \propto \Omega$), 
%than would be fit by the higher fluxes 
and (c) yield lower dust mass
determinations (from Equation~\ref{eqn:md}, $M_{\rm d} \propto S_{\nu}$)
than would be estimated with the higher submm 
fluxes.

\section{Decomposing the SED}\label{sec:sed_es}

To model the radio to infrared 
%spectral energy distribution 
SED, a
combination of a power-law  (non-thermal radio
emission from the galaxy's nucleus) plus greybody (re-processed emission
from dust) was used.  The composite model flux is 
\vspace{-0.15in}
\begin{equation}
F_{\nu} = C\nu^{\alpha} + \Omega B_{\nu}(T)
[1-{\rm exp}(-({\frac{\lambda_o}{\lambda}})^{\beta})],
\label{eqn:pb}
\end{equation}
%In equation (4.1),
where $C$ is the normalization for the power-law component,
$\alpha$ the power-law spectral slope with $F_{\nu} \propto {\nu}^{\alpha}$, $\Omega$ the solid angle for
the greybody component,  $B_{\nu}(T)$ the Planck function at
temperature $T$, $\lambda_o$ the wavelength at which the optical depth
is unity and $\beta$ the emissivity index of the grains.  
%the model in 
Due to the limited
frequency sampling in the data, the temperature $T$, and its uncertainty are
the only key parameters that are statistically determined in
Equation~\ref{eqn:pb}.  Therefore, in fitting Equation~\ref{eqn:pb},
the factors $C$ and $\Omega$ were normalized to the data, the
temperature was allowed to vary and statistically computed, and the
power-law index $\alpha$ and emission parameters $\lambda_o$ and 
$\beta$ were fixed and adopted, as described below.

The normalizing 
factors $C$ and $\Omega$
were obtained by 
%normalizing 
forcing the 
%equation
model flux to agree with the core-radio (1.4 and 5\,GHz) and FIR (60
and 100\,$\mu$m) fluxes 
respectively, since the two components dominate in these different 
wavebands; and, the best fitting values for 
the temperatures and their uncertainties
were determined by minimizing ${\chi}^{2}$ for preset values of $\beta
= 1, 1.5, 1.8$, and 2 in Equation~\ref{eqn:pb}.  Following
\citet{lee00} and references therein, $\lambda_o =7.9$\um\ and
$\alpha = -0.25$ were fixed and used on fits for all objects but
%.  The exceptions was 
%for 
NGC\,2986
and NGC\,5353, 
for which, from 
examinations of their 
fits by eye, the mm and submm data clearly required steeper power-law
slopes of $\alpha = -0.5$ 
%(cf. Table~\ref{tab:one}).  
and $-0.3$ respectively (cf. Table~\ref{tab:one}).  

The lowest ${\chi}^{2}$'s, i.e., best fits, were obtained
with $\beta = 2$, and the worst with $\beta = 1$.
In general, the fits  
became worse and yielded slightly higher temperatures with decreasing $\beta$.
Relative to values obtained with $\beta = 2$, the temperatures
obtained with $\beta = 1.8$ and $1.5$ were greater by
$T \sim 2$ and $\sim 4$\,K, respectively.    
Table~\ref{tab:one} lists the temperatures and ${\chi}^2$'s
for ${\beta} = 1.5$, as a demonstration of the higher temperatures and poorer
fits with $\beta < 2$.  
%This need not be overemphasized; because,
The table also shows that while values of $\beta = 2$ looks best, there is
not a lot to choose between that and values of $\beta = 1.5$,
indicating the degree of uncertainty in the fits. 
%Only NGC\,3962, NGC\,4374 and NGC\,5354 provide exceptions to this.
For 
%$T$ 
temperature values determined here, a choice of $\beta$ primarily affects
the SEDs longward of 100\um, so that 
%in the case of
for NGC\,3962 and NGC\,5354, for which
there are only flux upper limits or no FIR to submm data points 
longward of 100\um, the $\beta $
providing the best fit (i.e., lowest ${\chi}^2$) for these objects cannot be
discriminated.  
%'s for these objects are poorly
%constrained. 
%In the case of 
For NGC\,4374, the
differences in 60 and 100\um\ fluxes  
%for thisbobject 
are the dominant contributor to the ${\chi}^2$; and therefore, the
${\chi}^2$ with different $\beta$'s for this object are about the same.

Mindful of the fitting uncertainties, the general deduction in this
paper is that the reduced ${\chi}^2 $'s that are listed in
Table~\ref{tab:one} 
analysis 
demonstrate that single
temperature greybodies with $\beta=2$ produced reasonable
SEDs fits for 
the objects presented here (see also
Section~\ref{sec:2tsed} below).  For NGC\,3962 and
NGC\,4697 that appear to have better $\beta=1.5$, ${\chi}^2$ than
$\beta=2$, ${\chi}^2$, submm upper limits not included in the ${\chi}^2$
determination fall below the $\beta=1.5$ model plots but are
consistent with the $\beta=2$ plots.  
Therefore, greybodies with temperatures
obtained from these 
$\beta=2$ fits are adopted in this paper.  They are 
plotted as dotted lines in
Figure~\ref{fig:sed_aes} and their derived parameters listed in
Table~\ref{tab:one}.  The power-law plus greybody model is plotted
as solid or dashed lines (see
Figure~\ref{fig:sed_aes}), respectively for
power-law indices of NGC\,4374, that was determined from linear
regression fits by \citet{lee00},
and of the other galaxies, that have been fixed following the results
of NGC\,4374 as described above.  
New mm-submm observations 
obtained with the Nobeyama
Telescope and SCUBA are marked by open squares and circles respectively.  New FIR
data from $ISO$  
 are indicated by asterisks.  Other infrared 100 and 60$\mu$m data in
the figures are from $IRAS$, which together with all radio data were
obtained from NED. The downward arrows indicate 3-$\sigma$ upper limits. 

\subsection{Limitations of Data and SEDs for Individual Galaxies}
\label{sec:separate_sed}

Because of the submm and
mm faintness of sources presented and the limited sensitivity of
instruments used, about half of the submm and mm data points are upper limits.  
Therefore, some galaxies in the sample have little or incomplete
mm to FIR observational data sets to allow a model fitting
to the SEDs with the full range of variables that could be considered
as free parameters in Equation~\ref{eqn:pb}.  Also, some $ISO$ data
duplicate the $IRAS$ measurements at 60 and 100\um, such that, even when
the number of free parameters to fit to the data are
restricted as described
above (see Section~\ref{sec:sed_es}), a few of the galaxies are still left with too few data  
and too many free parameters in the fitting process.

Specifically, NGC\,2986 has only two photometric data points and three
upper limits in the mm to FIR regime.  
Therefore, no reduced ${\chi}^2$ was computed for this object.  The
dust temperature obtained can be thought of as derived from the 60 to
100\um\ flux ratio, as commonly derived for objects with only $IRAS$ data
\citep[e.g.][]{gou95}.  A tentative composite plot for this object is
displayed in Figure~\ref{fig:sed_aes}
%Figures~\ref{fig:sed1} 
to demonstrate the consistency of
the mm-submm upper limits and FIR data with the model fit.
Similarly, a tentative dust mass for this object is computed as
described for the other objects below and listed in Table~\ref{tab:one}.

It is noted that NGC\,3156 has limited number of data points.  However, the three
points this galaxy has in the FIR to submm region are sufficient for making the
required fit to the 
%two free parameters 
temperature and normalisation of its thermal spectrum.   It
is also noted that NGC\,4697 has only 3 upper limits in the cm-submm
region.  However, this object has a very good sampling of points around
the peak of the thermal spectrum; and so, for NGC\,4697, the upper
limits in mm-submm region 
are not needed or used to provide any constraints and are plotted
primarily to show that these
data
%upper limits 
are consistent with the fit. 

Currently only limited data is available for fitting the power-law
spectrum in the region from 1.4 to 5\,GHz.  
As noted earlier, the power-law index is therefore fixed at $-0.25$ for
all the galaxies but for NGC\,2986 and NGC\,5353, where it is estimated as $-0.50$
and $-0.30$ respectively.   
%As noted earlier, 
Fixing the index follows
fits of NGC\,4374 
in earlier work \citep{lee00}, since this source has the needed points from 1.4 to
5\,GHz.  Commissioning of high frequency radio instruments such as the
Green Bank Telescope should allow data in these wavebands to be
obtained for other ellipticals in the future, enabling more reliable
correction of non-thermal contribution to the mm-FIR thermal flux.
As noted earlier, the $C$ parameter normalizes the
power-law component of the model to the 1.4 and 5\,GHz data, for either
detected points or upper limits, depending on the presently scant
availability of needed data in literature and archives.  This
normalization is done well
aware that it is less reliable in the cases such as for NGC\,3156 and
NGC\,4697, where only 5\,GHz upper limits are currently available.

\subsection{
A Second Dust Component with $T < 20$\,K?}\label{sec:2tsed}

%Investigations
Spectral energy distribution analyses by 
%Dunne \& Eales 2001 
\citet[][]{dun01} and \citet[][]{kla01}, 
%Klaas et al 2001, 
among
others, have demonstrated evidence for thermal emission with at
least two temperature components and an emissivity index of 2.  This
was for samples of bright $IRAS$ and ultra-luminous infrared
galaxies and was deduced not 
only from rigorous fitting results but also physical considerations.  
The presence of very cold dust has also been 
%observed 
suggested
 in some
individual sources, e.g. the peculiar dwarf elliptical galaxy NGC 205
\citep[][]{haa98}
%(Haas 1998, A \&A 337, L1) 
and the merger-remnant elliptical galaxy
%in the merger-remnant elliptical 
NGC\,5128 \citep[]{lee02}.   

\begin{figure}
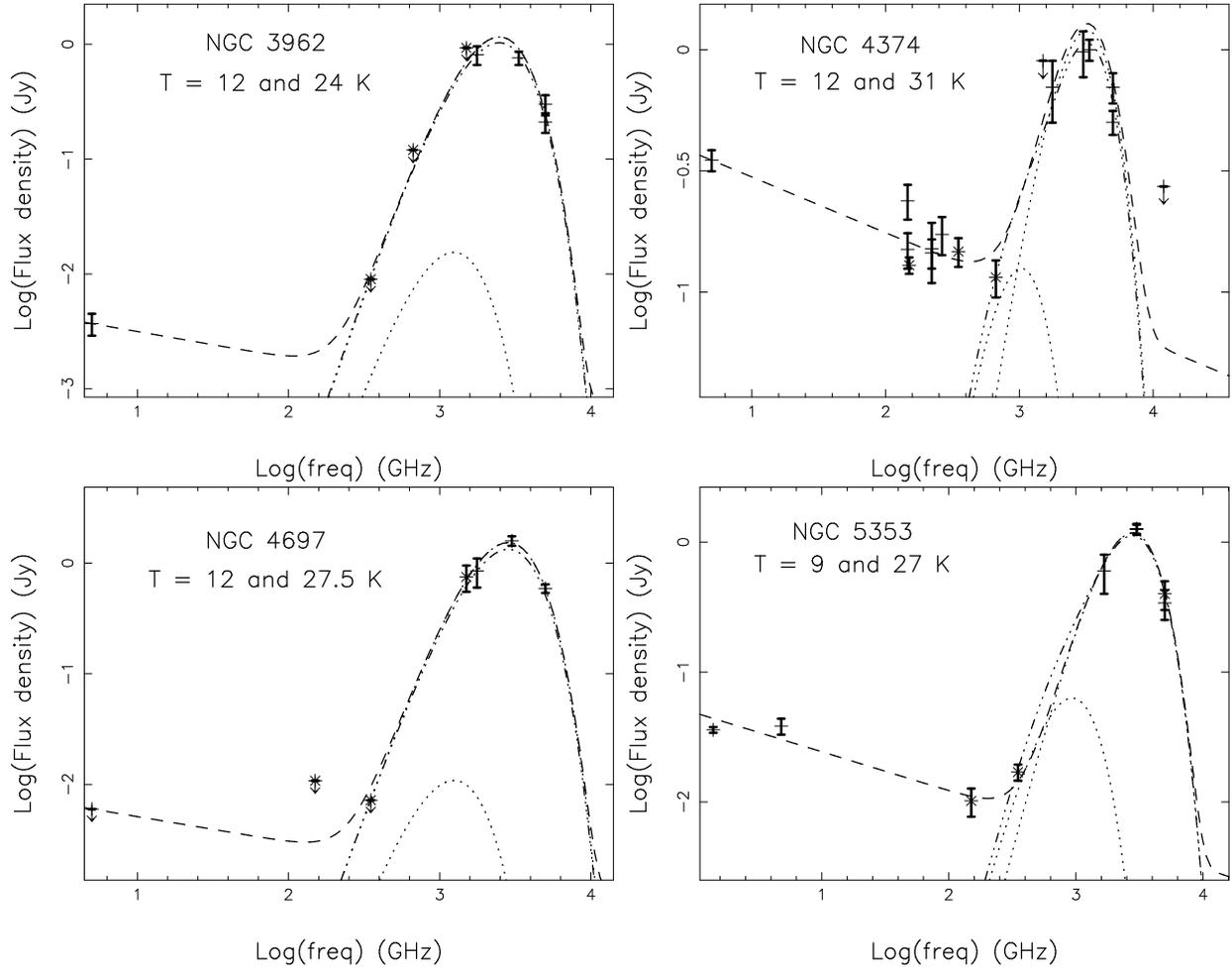

%\plottwo
\begin{center}
\includegraphics[angle=-90,width=3.2in]{f3a.eps}
\includegraphics[angle=-90,width=3.2in]{f3b.eps} \\
\includegraphics[angle=-90,width=3.2in]{f3c.eps}
\includegraphics[angle=-90,width=3.2in]{f3d.eps}
\end{center}
\vspace{-0.1in}
\caption{Radio to far-infrared spectral energy distributions 
%Spectral energy distributions from radio to infrared wavelengths
of central regions in sample FIR-bright elliptical galaxies. 
%NGC\,2986, 3156, 3962, 4374, 4697 and 5353.  
The model fits are of a two-temperature greybody given by
Equation~\ref{eqn:2tsed} and they together with the data symbols are as
described in the text.
\label{fig:2tsed}}
\end{figure}

Following
the investigations by 
%Dunne \& Eales 2001 
\citet[][]{dun01} and \citet[][]{kla01}, 
the 
%presented 
FIR to submm data are also fit with
two-component greybodies 
%and exploited 
to 
%constrain examine physical parameters for
test whether 
%two-component 
%two-temperature
such greybodies,
%the emission 
%particularly 
including those with components of temperatures less than 20\,K (see
Figure~\ref{fig:2tsed}), produce better fits than single-temperature models.
%greybodies.   
%Though limited 
%by the uncertainty in
%by the available data points, 
%The fact 
That the single-temperature greybodies alone do not fit both FIR and
submm data simultaneously may indicate that there is a second very
cold (i.e., $<20$\,K) dust component required.  
The data longward of
100\um, as newly available for this paper, can sample fluxes from dust
with temperatures not only in the $IRAS$ range of $T\sim 20$ to
100\,K, but also less than 20\,K.   
However, the current
data coverage is too sparse in wavelength space and in some cases 
the detections have relatively large uncertainties or only flux upper
limits are available.  Therefore, it is not possible to {\em reliably}
%only tentative 
determine the exact
decompositions of the 6 to 20\,K (very cold) and 20 to 40\,K (cold)
components   
%{\em reliably}
or {\em securely}
discriminate 
whether two-temperature fits are better than single-temperature ones
for all galaxies.
%are presently possible.    
%Therefore, 
%However

For now, the SEDs  for which there were at least three data points
%available 
from 100 to 850\um\ 
%were also fitted with  do 
allow an eyeball fit
that is useful for
estimating the flux upper limit for the very cold dust temperatures component
that is 
consistent with the data.  
In this regard and to facilitate comparison with fits from Equation~\ref{eqn:pb}, a two-temperature component greybody of
the form
%\vspace{-0.15in}
\begin{equation}
F_{\nu} = C\nu^{\alpha} + \Omega [B_{\nu}(T_1) + B_{\nu}(T_2)][1-{\rm
exp}(-({\frac{\lambda_o}{\lambda}})^{\beta})],
\label{eqn:2tsed}
\end{equation}
where the parameters are as in Equation~\ref{eqn:pb}, was fit to the
FIR-submm data of NGC\,3962, NGC\,4374, NGC\,4697 and NGC\,5353
(see Figure~\ref{fig:2tsed}). 
The modelled-flux sums of 
two-temperature greybodies are normalized to detected-FIR or -submm
fluxes, and it is assummed that the greybodies for respective galaxies have
the same solid angles and emissivity indices as obtained from
Equation~\ref{eqn:pb} and listed in Table~\ref{tab:one}.  

Figure~\ref{fig:2tsed} exhibits the composite
%greybodies 
and individual greybodies that constitute the composites
 by dashed-three-dotted and purely dotted lines, respectively.
%The individual greybody curves fall below the submm data,
Because it
is the sum of the two components that are normalized to the data,
curves of the composite and individual greybodies respectively fall
on and below the submm data.  For NGC\,3962, the observed 850\um,
3-$\sigma$ limit of 0.009\,Jy is fit by 12 and 24\,K greybodies with
modelled 850\um\ 
fluxes of 0.0012 and 0.0078\,Jy, respectively; for NGC\,4374, the
observed 450\um\ flux
of 0.110\,Jy is fit by 12 and 31\,K greybodies with modelled 450\um\
fluxes of 0.086 and 0.034\,Jy, respectively; for NGC\,4697, the
observed 850\um, 3-$\sigma$ limit
of 0.007\,Jy is fit by 12 and 27.5\,K greybodies with modelled 850\um\
fluxes of 0.009 and 0.0061\,Jy, respectively; and, for
NGC\,5353, the observed 850\um\ flux
of 0.017\,Jy is fit by 9 and 27\,K greybodies with modelled 850\um\
fluxes of 0.011 and 0.006\,Jy, respectively.  The modelled-submm fluxes
of very cold dust components provide 
flux estimates
%upper limits 
of emission that
would originate from a second dust component at the respective
temperatures
and may be consistent with the submm-FIR data.
If there's any 
%non-thermal 
additional flux contribution to observed submm flux, the
modelled flux from very cold dust components will scale down in
proportion to the contributing flux.  

Although the two-temperature
model may be consistent with the data, it produced relatively high
reduced ${\chi}^{2}$ of 
%\ldots,
2, 4.6, 1.6, 2.2 
respectively for NGC\,3962, NGC\,4374, NGC\,4697, and NGC\,5353
(see Table~\ref{tab:one}), 
showing that under  the fitting criteria
assumed in this paper, the single-temperature model
%greybodies 
of the form in Equation~\ref{eqn:pb} fits the presented data slightly
better than the two-temperature model.
Following this comparison and the fact that the 
%presented 
model fits 
are based on under-sampled SEDs with many upper limits, 
%these analysis
%cannot positive
this paper 
cannot positively identify (or rule-out) the presence of cold dust
%less than 
$<20$\,K in the presented
%sample 
elliptical galaxies.

\section{Derived Dust Parameters 
%and their Implications
}\label{sec:para_es}  

The best fitting power-law plus greybody
model to the radio-to-infrared SEDs constrains the dust to central
galactic regions of diameters $\sim
0.4$ to 2.4\,kpc ($5''$ to $9''$), average dust temperatures between
$\sim 21$
and 28\,K, and emissivity index $\beta \simeq 2$ (see Table~\ref{tab:one}).
%$\pm 2$\,K 
%, or specifically $\beta = 1.9 \pm 0.2$.  
It was noted in Section~\ref{sec:sed_es} that
when the new submm to $ISO$-FIR data are not considered, such as in
the ${\chi}^2$ determinations that do not use the submm-FIR upper
limits, it appears that some model fits with $\beta =1.5$ are as good or
better than those with  $\beta =2$ (see Table~\ref{tab:one}).
However, it was also noted that for NGC\,3962, NGC4374, and
NGC\,4697, submm upper limits not included in the ${\chi}^2$
determination fall below the $\beta=1.5$ model plots but are
consistent with $\beta=2$ plots, favoring models with $\beta=2$
(cf. Section~\ref{sec:sed_es}).     
Because the size of a dust emitter is inversely propotional to the
dust emissivity index and temperature, the diameters derived for
$\beta =2$ are upper limits for $\beta$ that ranges for 1 to 2 and may
be larger for dust with temperatures lower than plotted in
Figure~\ref{fig:sed_aes}.  However, such larger emitting regions would
be required to emit flux levels much lower than detected in the presented
observations. 
%.. requires that the colder temperature have very lower fluxes
%andto-fill  diameters n.
 The dust temperatures
around $20$\,K and and emissivity index ($\beta \approx 2$),
especially, show that dust in ellipticals can exhibit properties
similar to those in the Galaxy (cf. dust temperature of 16 to 21\,K and
emissivity index of 2, \citet{rea95_et}), and probably constitute similar
model-predicted amorphous silicate and carbonaceous grains \citep{li01}.
%Coincidentally, these 
%that are by dust models for dust there 
%are probely similar.  

Following 
%work by 
\citet[][]{hil83}, the mass of emitting dust $M_{\rm d}$ can then be estimated from
\begin{equation}
M_{\rm d}= \frac{{S_{\nu} D^2}}{{k_{\rm d} B(\nu ,T)}},
\label{eqn:md}
\end{equation}
\noindent
where $S_{\nu}$ is the measured flux density at frequency $\nu$, $D$ is the
distance to the source, $B({\nu} ,T)$ the Planck function and $k_{\rm d} =
3Q_{\nu}/4a\rho$ the grain mass absorption coefficient where $a$ and
$\rho$ are respectively the grain radius and density.  Recently
updated values of
$k_{\rm d}^{100\,\mu {\rm m}} = 3.5\,{\rm m}^2 {\rm kg}^{-1}$
and $k_{\rm d}^{60\,\mu {\rm m}} = 10.62\,{\rm m}^2 {\rm kg}^{-1}$
\citep[][]{li01}, for Galactic-like dust (cf. above),
%\citep{li01} 
are assumed, yielding dust masses that range
from $\sim 1.6
\times 10^5 {\rm M}_{\odot}$ to $1.9 \times 10^6 {\rm M}_{\odot}$ (see
Table~\ref{tab:one}).

These
% presented
dust masses are of similar range as determined in the FIR studies by
\citet[][]{rob91} and \citet[][]{gou95}, even though those workers
used $k_{\rm d}$ values that are typical for dust with emissivity index less
than 1.5 \citep[e.g.][]{hil83,dra90} and about 3 times less than the
values used here.  The
difference in the employed $k_{\rm d}$ values is balanced by the fact that 
dust temperatures derived here are slightly cooler than when estimated
from FIR data alone \citep[c.f.][]{rob91, gou95, bre98}, leading to
lower values of $B(\nu ,T)$ for this work and the similarity in the dust mass
estimates.  One advantage of the current study is that, instead of
assuming the dust emissivity index and
estimating the dust temperatures from the $IRAS$ 60\um\ and 100\um\ flux ratio
\citep[c.f.][]{rob91, gou95}, the emissivity index and dust temperatures have 
been constrained by 
%derived from 
new FIR and submm to mm data that extend to longer
wavelengths than earlier $IRAS$ observations and better sample the
Rayleigh-Jeans tail of the cold dust thermal spectrum.  As noted earlier,
the resulting constraints 
%strongly 
suggest that dust in ellipticals has
%very 
similar properties as Galactic dust.   

Derived parameters for NGC\,2986 have 
relatively high uncertainties 
that are propagated 
from the $IRAS$ and SCUBA data.  
$IRAS$ data that were re-processed with HIRES and
SCANPI routines at the NASA/IPAC Center
indicate that this source's $IRAS$ flux is probably contaminated by a
foreground or background object.  

For the two galaxies NGC\,3962 and NGC\,4374, for which the respective
$10^{4.66} {\rm M}_{\odot}$ and $10^{4.54} {\rm M}_{\odot}$ dust masses
estimated from optical-extinction studies are available \citep{gou95}, the dust content
determined here is about a magnitude
larger than that determined from the optical studies of
ellipticals, as observed by \citet{gou95} and \citet{sil96a}.  Both
authors proposed that a diffusely distributed dust
component not properly accounted for in the optical studies may be
responsible for the extra dust mass determined in the infrared.  The
diffuse dust would probably be heated by a stellar radiation field
that is more dilute than in the central regions and thus be relatively
cold.  This is withstanding the fact the dust can also
be heated by the collision with electrons in 
the hot gas, with a contribution of the same order of the photon
heating even at large radii \citep[][]{tem03}.
%(Temi et al. 2003, ApJ, 585). 
  The data presented here are consistent with emission from dust
marginally warmer than that 
seen in some FIR-bright, nearby spiral galaxies \citep[e.g.][]{dun01},
and allow flux and dust-mass 
%upper limits 
that are consistent with the data to be estimated for 
%may show evidence of emission from 
dust cooler than 20\,K (see below) within inner galactic regions.

\subsection{Masses for Dust with $T < 20$\,K}\label{sec:2tpara_es}  

As above, the mass
$M_{\rm d}$ of FIR-to-submm emitting dust with temperature $< 20$\,K 
%that emits radiation
%can then be 
is 
%estimated
determined using Equation~\ref{eqn:md}; and, for the two-temperature
dust fitting results, assuming 
%Recently updated 
values of
$k_{\rm d}^{450\,\mu {\rm m}} = 0.157\,{\rm m}^2 {\rm kg}^{-1}$
and $k_{\rm d}^{850\,\mu {\rm m}} = 0.050\,{\rm m}^2 {\rm kg}^{-1}$
\citep[][]{li01}.
%, for Galactic-like dust. 
%(cf. above),
%\citep{li01} 
%are assumed,   
For (a) NGC\,5353, T=9\,K and $S_{\nu}^{850\,\mu {\rm m}} = 0.011$\,Jy
yield $1.61 \times 10^4 {\rm M}_{\odot}$; (b) NGC\,4697, T=12\,K and $S_{\nu}^{850\,\mu {\rm m}} = 0.00087$\,Jy
yield $8.31 \times 10^4 {\rm M}_{\odot}$; (c) NGC\,4374, T=12\,K and $S_{\nu}^{450\,\mu {\rm m}} = 0.086$\,Jy
yield $6.87 \times 10^5 {\rm M}_{\odot}$; and (d) NGC\,3962, T=12\,K and $S_{\nu}^{850\,\mu {\rm m}} = 0.0012$\,Jy
yield $2.71 \times 10^5 {\rm M}_{\odot}$. 
% As t
The fluxes used assume
the only contribution to the detections is
%se fluxes are 
the ``very cold'' and ``cold'' dust, 
%i.e. not non-thermal (
which may not be the case.
%), these masses are upper limits for ``very cold'' dust.

The 12\,K dust mass estimate for NGC\,4374 is of the same order as the
warmer 28\,K dust mass determined for this galaxy.  Therefore, if the
$450\,\mu {\rm m}$ 
%submm
flux for this object is primarily from very cold dust, this component
could be as significant as that from the 28\,K dust.  For this to be the case,
the high-frequency, non-thermal power law would have to 
%significantly dust mass upper limitdust mass upper limit
turn-over or cut-off before the 
%submm
$450\,\mu {\rm m}$ flux.  The 12\,K dust mass estimate for NGC\,3962 is also of the same order as the
warmer 24\,K dust mass determined for this galaxy.  In this case, this
is all that can be said because only upper limit fluxes are available
in the submm and no radio observations have been made to constrain any
low frequency flux origin for this galaxy. 
For NGC\,5353 and NGC\,4697, the 9\,K dust mass estimate is about two orders of
magnitude less than the warmer 28\,K and the 12\,K dust mass estimate
about an order of 
magnitude less than the warmer 25\,K dust masses determined for these
galaxies, respectively.  Therefore in these cases, the cooler
components, though of relatively considerable mass on their own, do not appear
to be as significant as the warmer components within the central
galactic regions.

\subsection{Implications of the Fitted Dust Parameters}\label{sec:impara_es}  

It is very interesting that the best fitting greybody model
parameters require dust emitting 
in the mm to FIR to be centrally concentrated and less
extended than the stellar component of ellipticals.  Because red giant
stars lose dusty gas to the ISM, if dust in
ellipticals derives from these stars, the dust and stellar spatial
distributions would be expected to be co-spatial.  
%, and dust may
However, assuming that the dust is indeed produced by stars, as gas  
cools down and falls into the potential well,
distributed dust may drift toward the galactic centers and settle into
a central disc.  Models of centrally concentrated, distributed dust and 
the transfer of stellar radiation in ellipticals
by \citet{wit92}
show that dust in such geometry 
would not affect
the observed $r^{(1/4)}$ surface brightness law (the de Vauculeaurs
profile) caused by the stellar radiation.   Moreover, such a distribution of dust 
would produce and naturally explain the
central reddening (or a color gradient that changes with radius) that
is observed in elliptical galaxies and
cannot be completely explained by models that consider only evolutionary stellar
population synthesis and metallicities \citep[e.g.][]{pel90}.

Recent dynamical modeling of dusty
gas ejected from evolved stars in ellipticals (or cD galaxies)
supports a stellar origin of the central dust clouds seen in these
galaxies \citep{mat03}.  The  models show that, even after entering the
hot X-ray emitting gas, rapid cooling by thermal collisions with dust
grains, especially in the central regions, can be faster than the
dynamical time in the galactic potential or grain sputtering time,
ensuring that some dust survives even in
ellipticals 
with X-ray emitting gas.  
Observations indeed show that some dust exists in ellipticals; 
and, as noted by \citet[]{kna99}, 
the dust probably 
suggests ellipticals host modest star formation.

\section{Conclusions}\label{sec:con_es}  

The current observations have detected centrally concentrated dust in
ellipticals with properties similar to those in the Galaxy
\citep[e.g.][]{rea95_et} and other
nearby FIR-bright galaxies \citep[e.g.][]{dun01}.
%spirals of similar
%These estimates imply 
Single-temperature, greybody-fitting results imply the dust is slightly cooler 
%and thus more massive than 
than when estimated from FIR data alone; however, dust masses 
determined using updated dust absorption
coefficients \citep[][]{li01} for Galactic-type dust yield masses
similar to estimates
from the FIR studies\citep[e.g.][]{rob91, gou95}.  
Tentative determinations have been made of 
flux and mass estimates for dust cooler than 20\,K, within the central
galactic regions.  Resolved submm-FIR observations show
a galactic dust temperature decrease from the center to the outer
regions \citep[e.g.][]{lee02},
where the radiation field heating the dust is dilute; and in
ellipticals, the colder dust is most probably 
cool, diffusely distributed and heated by both the dilute stellar
radiation and
the collision with electrons in the hot gas \citep[][]{tem03}.  Because of 
the general paucity of the dust and the observation that it may be central
concentrated (see above), the direct
detection of diffuse, extended dust in 
ellipticals is challenging.  
If such   
diffuse dust with {\it low} surface brightness exists, observationally
constraining its physical properties 
%properties further further for
%will have to wait for 
may be possible with deeper SCUBA observations (especially mapping) or
soon to be commissioned or built sensitive instruments
like the {\it Spitzer Space Telescope}, SCUBA2, and ALMA.   

\acknowledgments

%\section{Acknowledgments}

Lerothodi L. Leeuw 
%(LLL) 
is supported by a NASA Grant NAG5-9376.  
Martin Haas is supported by the Nordrhein-Westf\"alische Akademie der
Wissenschaften,
funded by the Federal State and the Federal Republic of Germany.

\clearpage

%\bibliographystyle{apj}
%\bibliography{lero_bib}

\clearpage

\begin{figure}
%\plottwo
\begin{center}
\vspace{7in}
\end{center}
\vspace{-0.1in}
\caption{The infrared Digital Sky Survey (DSS) 2 images of NGC\,5353 and
NGC\,5354 with superposed ISOPHOT C100 (top-left panel) and C200
(top-right panel) array pointings that were used for the $ISO$
observations of these galaxies.  The bottom-left and -right panels
show 3-D histograms of the $ISO$ 100 and 180\,\um\ fluxes, respectively,
before the individual-galaxy fluxes were corrected for aperture size
(see text).  For visual clarity, the 
histograms are oriented so that the fainter source is plotted in
the foreground. \label{fig:arrays}} 
\end{figure}

\clearpage

\begin{figure}
%\plottwo
\vspace{7in}
\begin{center}
\vspace{-0.9in}
%\includegraphics[]{f2.eps} \\
%angle=-90,width=2.5in
\end{center}
\vspace{-0.1in}
\caption{Radio to far-infrared spectral energy distributions 
%Spectral energy distributions from radio to infrared wavelengths
of central regions in sample FIR-bright elliptical galaxies. 
%NGC\,2986, 3156, 3962, 4374, 4697 and 5353.  
The model fits include a composite power law plus greybody given by
Equation~\ref{eqn:pb} and they together with the data symbols are as described
in the text. 
\label{fig:sed_aes}}
\end{figure}

\clearpage

\clearpage

%\begin{table}[hbt!]
\begin{deluxetable}{lccccccccc}
%\hline
\tabletypesize{\scriptsize}
\tablecaption{
Far-infrared to mm fluxes
of nearby $IRAS$-bright ellipticals. \label{tab:zero}} 
%\hline
%Errors are one sigma.} 
\tablewidth{0pt}
%\vspace{0.2cm}
%\begin{tabular}{|l|c|c|c|c|c|c|} \hline
\tablehead{
\colhead{Galaxy} & \colhead{$IRAS$} & \colhead{$IRAS$} &
\colhead{$ISO$} & \colhead{$ISO$} & \colhead{$ISO$} & \colhead{$ISO$} & \colhead{SCUBA} & \colhead{SCUBA} & \colhead{NOBA} \\
%Galaxy	& $IRAS$ & $IRAS$ & $ISO$ & $ISO$ & SCUBA & SCUBA & NOBA \\
%Name	& 60\um & 100\um & 170\um & 200\um & 450\um & 850\um & 2\,mm \\
\colhead{Name}	& \colhead{60\um} & \colhead{100\um} &
\colhead{60\um} & \colhead{90\um} & \colhead{170\um} &
\colhead{200\um} &  \colhead{450\um} & \colhead{850\um} & \colhead{2\,mm}
%\\ \hline
}
\startdata
NGC\,2986 & $< 0.05$ & $0.36 \pm 0.11$ & \nodata & \nodata & \nodata & \nodata & $<
0.13$ & $< .010$ & \nodata \\    
NGC\,3156 & $0.22 \pm 0.04$ & $0.91 \pm  0.09$ & \nodata & \nodata & \nodata & \nodata
& $<0.06$ & $.006 \pm .002$ & \nodata \\ 
NGC\,3962 & $0.21 \pm 0.04$ & $<0.98$ & $0.30 \pm 0.06 $ & $0.76 \pm
0.10$ & $0.81 \pm 0.15 $ & $ < 0.91 $  &
$<0.12$ & $ <.009$ & \nodata \\  
NGC\,4374 & $ 0.50 \pm 0.06 $ & $ 0.98 \pm 0.21 $ & $ 0.70 \pm 0.10 $ & $ 1.00 \pm 0.10 $
& $ 0.70 \pm 0.20 $ 
& $ < 0.90 $ & $ 0.12 \pm 0.02 $ & $ .147 \pm .020 $ & $ .129 \pm .010 $ \\
NGC\,4697 & $ 0.59 \pm 0.05 $ & $ 1.24 \pm 0.08$ & $1.15 \pm 0.35$\tablenotemark{a} & $ 1.45
\pm 0.35 $\tablenotemark{a} & $  0.85 \pm 0.25 $ & $  0.75 \pm 0.20 $ & \nodata & $ < .007  $ & $ < .010 $ \\
NGC\,5353 & $ 0.32 \pm 0.06 $ & $ 1.45 \pm 0.11 $  & 
$ 0.40 \pm 0.10 $  & $ 1.25 \pm 0.10 $\tablenotemark{b}  & $ 0.60 \pm 0.20 $\tablenotemark{c}
%\nodata & \nodata & \nodata
& \nodata & \nodata & $ .017 \pm
.002 $ & $ .011 \pm .003 $ \\  
NGC\,5354 & $ 0.41 \pm 0.05 $ & $ 1.61 \pm 0.06 $ & 
 $0.20 \pm 0.06$ & $0.70 \pm 0.10$\tablenotemark{b} & $0.58 \pm 0.20$\tablenotemark{c} &
\nodata & \nodata & \nodata  & \nodata \\  
%\hline
 \enddata

%% Text for table notes should follow after the \enddata but before
%% the \end{deluxetable}. Make sure there is at least one \tablenotemark
%% in the table for each \tablenotetext.

\tablenotetext{a}{There is a large uncertainty in the on-board
calibration measurement for this flux.}
\tablenotetext{b}{This is an $ISO$ 100\um\ filter measurement from
the C100 array.}
\tablenotetext{c}{This is an $ISO$ 180\um\ filter measurement from
the C200 array.}

\tablecomments{All listed fluxes are in Jansky (Jy), and errors and
upper limits are respectively 1 and 3\,sigma.}

%\end{tabular}
%\end{table}

\end{deluxetable}

\clearpage

\begin{deluxetable}{lcccccccccccc}
%\hline
\tabletypesize{\scriptsize}
\tablecaption{Power-law plus greybody 
%model 
fitting results 
%FIR to short-radio photometry 
for nearby, $IRAS$-bright ellipticals.  \label{tab:one}} 
\tablewidth{0pt}
\tablehead{
\colhead{Galaxy} & \colhead{Galaxy} & \colhead{Effect.} &
\colhead{D} & \colhead{P-L\tablenotemark{a}}  & \colhead{$T_{\rm
dust}$} &
\colhead{$\Omega_{\rm dust}$} & 
%\colhead{} &
%\colhead{${\chi}^{2}$} &
%\colhead{No. of} &
\colhead{Red.} &
%\colhead{${\chi}^{2}$} &
%\colhead{$\chi^{2}_{\rm fit}$} &
%^{170\,\mu{\rm m}}$} &
\colhead{M$_{\rm dust}^{60\,\mu{\rm m}}$} 
%& \colhead{M$_{\rm dust}^{\rm{opt.}}$} 
& \colhead{$T_{{\rm dust}}$\tablenotemark{b}} 
& \colhead{Red\tablenotemark{b}}
& \colhead{$T_{{\rm dust}}$\tablenotemark{c}} 
& \colhead{Red\tablenotemark{c}}
\\
\colhead{Name}	& \colhead{Type}  & \colhead{Radii} & \colhead{Mpc} &
\colhead{index}     & \colhead{(K)} 
& \colhead{(ster.)} & 
%\colhead{} &
%\colhead{${\chi}^{2}$} &
%\colhead{Points} &
\colhead{${\chi}^{2}$} 
%& \colhead{\nodata}
%($log({\rm M}_{\odot})$)} 
& \colhead{($log({\rm M}_{\odot})$)} 
%& \colhead{($log({\rm M}_{\odot})$)}
& \colhead{(K)}
& \colhead{${\chi}^{2}$} 
& \colhead{(K)}
& \colhead{${\chi}^{2}$} 
%& \colhead{}
}
\startdata
NGC\,2986	& E2 & $ 36''$ & 40.2  & $-0.50$	& $\sim 21 $ ($<32$) & 
(1.04E-9)\tablenotemark{d} & 
%N/A &	
%&  \nodata 
%1 &
N/A & ($\sim 6.28$)\tablenotemark{d} 
%& \nodata \\
&  $\sim 23$ &  N/A 
&  N/A &  N/A \\
NGC\,3156	& E/SO & $ 14'' $ & 22.3 & $-0.25$	& $24.6 \pm 0.9 $
&  1.26E-9 & 
%0.23 &
%& \nodata 
%3 & 
0.1 & $5.60 \pm .02$ 
%& \nodata  \\
& 28.7 & 3.8
%&  7.68 
&  N/A &  N/A \\
NGC\,3962	& E1 & $35'' $ & 32.4 & $-0.25$ & $24.0 \pm 0.7$ &  1.67E-9 &
% 3.92 & 	
%& \nodata 
%4 & 
1.3 & $5.67 \pm .02$ 
%& 4.66 \\
&  26.2 &  
%3.55 &
1.2 & 12/24 & 2.0 \\
NGC\,4374	& E1 & $51'' $ & 20.7 & $-0.25$ & $28.1 \pm 0.9$ & 0.84E-9 &
% 29.9	&
%& \nodata 
% 13 & 
2.5 & $5.20 \pm .02$ 
%& 4.54 \\
& 31.5  
%&  29.9
& 2.8 & 12/31 & 4.6 \\
NGC\,4697	& E6 & $72'' $ & 21.4 & $-0.25$	& $28.1 \pm 0.7$ & 0.94E-9 &
% 6.93 &	 
%& \nodata 
%6 & 
1.4 & $5.60 \pm .02$ 
%& \nodata \\
&  31.2 
%&  4.16 
& 0.8 & 12/28 & 1.6 \\
NGC\,5353	& E/SO & $15'' $ & 45.1 & $-0.30$  & $25.3 \pm 0.9$ & 1.46E-9
& 
%6.34 &
%& \nodata 
%7 & 
1.0 & $6.15 \pm .02$ 
%& \nodata \\
&  29.2 
%&  6.86
& 2.5 & 9/27 & 2.2 \\
NGC\,5354	& SAO: sp & $18'' $ & 45.1 & $-0.25$  & $24.6 \pm 1.8$
& 9.00E-10 & 
%0.21 &
%& \nodata 
%3 & 
0.1 & $6.16 \pm .04$ 
%& \nodata 
&  26.9 
%&  0.02
& 0.01 & N/A & N/A  
\enddata

%% Text for table notes should follow after the \enddata but before
%% the \end{deluxetable}. Make sure there is at least one \tablenotemark
%% in the table for each \tablenotetext.

\tablenotetext{a}{The adopted power-law slope of $-0.25 \pm 0.3$ for NGC\,4374
  is that previously fitted to the radio-to-submm data by
  \citet{lee00}.  As only limited radio to submm
  data are available for the other objects have, the other indices are
  fixed as close as the fitting reasonably permits to the value for NGC\,4374.
}

\tablenotetext{b}{These values are for $\beta=1.5$, for comparison to
the values for $\beta=2$ adopted in this paper.  For NGC\,3962 and
NGC\,4697 that have better $\beta=1.5$, ${\chi}^2$ than
$\beta=2$, ${\chi}^2$, submm upper limits not included in the ${\chi}^2$
determination fall below the $\beta=1.5$ model plots and favor the
$\beta=2$ model.}

\tablenotetext{c}{These values are for two-temperature components with
$\beta=2.0$, for comparison to the values for single-temperature model fits.}
%in the rest of the table.}

\tablenotetext{d}{The parenthesis indicate a large uncertainty due to
limited data, and the dust mass 
%for this galaxy 
is calculated from the
100\um\ flux.}
%calibration measurement for this flux.}
%\tablenotetext{b}{This is an $ISO$ 100\um\ filter measurement from
%the C100 array.}
%\tablenotetext{c}{This is an $ISO$ 180\um\ filter measurement from
%the C200 array.}

\tablecomments{The galaxy types and effective radii were respectively
%optical diameters in arcminutes were 
obtained from NED and \citet[][]{dev92}(RC3).  The distances 
%and optically-estimated dust masses listed respectively 
listed in column four
%and nine 
are from earlier FIR studies \citep{rob91, gou95} to
best facilitate comparison.  The listed uncertainties are 1 sigma.}

%\end{tabular}
%\end{table}

\end{deluxetable}

\end{document}